\documentstyle[aps]{revtex}

\begin{document}
\draft
\title{Note on the derivative of the hyperbolic cotangent}
\author{G. W. Ford}
\address{Department of Physics, University of Michigan\\
Ann Arbor, MI 48109-1120}
\author{R. F. O'Connell}
\address{Department of Physics and Astronomy, Louisiana State University,\\
Baton Rouge, LA 70803-4001}
\maketitle

\begin{abstract}
In a letter to Nature (Ford G W and O'Connell R F 1996 Nature 380 113) we presented a formula for the derivative of the hyperbolic cotangent that differs from the standard one in the literature by an additional term proportional to the Dirac delta function. Since our letter was necessarily brief, shortly after its appearance we prepared a more extensive unpublished note giving a detailed explanation of our argument. Since this note has been referenced in a recent article (Estrada R and Fulling S A 2002 J. Phys. A: Math. Gen. 35 3079) we think it appropriate that it now appear in print. We have made no alteration to the original note. \end{abstract}

In Ref. \cite{ford96} we published the formula 
\begin{equation}
{\frac{d}{dx}}\coth (x)=-\text{csch}^{2}(x)+2\delta (x),  \label{1}
\end{equation}
and gave an argument showing that it is correct. In this note we give some
additional detail on the derivation of the formula. First, however, it might
be useful to point out that the function $\coth (x)$ increases by $+2$ as $x$
goes from $-\infty $ to $+\infty $. Yet its derivative is everywhere
negative, except at $x=0$. How can a function that is everywhere decreasing
still increase? We shall see how the answer is given by this formula.

We should emphasize that, as should be obvious from the appearance of the
Dirac delta-function, this is a formula of {\em distributions}. As a
function, $\coth (x)$ and its derivative are undefined at $x=0$, but as
distributions they can be given meaning for all real $x$ and it is for these
distributions that the formula is correct. In general a distribution is the
limit of a sequence of good functions \cite{light}, where a good function
and all its derivatives are continuous and bounded for all $x$. We in the
following give an explicit example of such a sequence for the various terms
in the formula.

We can define $\coth (x)$ as a distribution as the limit as $\epsilon
\rightarrow 0$ of the good function, 
\begin{equation}
\coth (x,\epsilon )\equiv {\rm Re}\{\coth (x+i\epsilon )\}={\frac{\sinh (2x)%
}{\cosh (2x)-\cos (2\epsilon )}}.  \label{2}
\end{equation}
For small $\epsilon $, this function is very close to $\coth (x)$ except in
a narrow range about $x=0$ where, instead of diverging, it turns over and
smoothly connects through the origin. This is shown in Fig. 1 for $\epsilon
=0.05$.

The derivative of this function is also a good function, 
\begin{equation}
{\frac{d\coth (x,\epsilon )}{dx}}=2{\frac{1-\cos (2\epsilon )\cosh (2x)}{%
[\cosh (2x)-\cos (2\epsilon )]^{2}}}.  \label{3}
\end{equation}
If we plot this function, we see that for $\epsilon $ small it will be very
close to $-$csch$^{2}(x)$ except for a narrow range of width of order $%
\epsilon $ about $x=0$, where there is a large positive peak. This is shown
in Fig. 2. The area under the central peak must exceed the (negative) area
under the wings by exactly $+2$, since that is the net change of $\coth
(x,\epsilon )$ as $x$ is carried from $-\infty $ to $+\infty $. This is
exactly accounted for by the delta-function in the formula (1). Thus, the
term $-$csch$^{2}(x)$ in that equation is to be considered as a distribution
with area zero \cite{note1}.

As a more explicit and detailed example of this separation, we write the
right hand side of (2) as the sum of two good functions, the first of which
has zero net change as $x$ is carried from $-\infty $ to $+\infty $, while
the second will have a net change of two and its derivative will approximate
the delta function. Thus, we can write (2) in the form 
\begin{equation}
\coth (x,\epsilon )=F(x,\epsilon )+G(x,\epsilon ),  \label{4}
\end{equation}
where $F$ and $G$ are good functions given by 
\begin{eqnarray}
F(x,\epsilon ) &=&{\frac{\sinh (2x)}{\cosh (2x)-\cos (2\epsilon )}}-{\frac{1%
}{\pi /2-\epsilon }}\arcsin ({\frac{\cos (\epsilon )\sinh (x)}{\sqrt{\sinh
^{2}(x)+\sin ^{2}(\epsilon )}}}),  \nonumber \\
G(x,\epsilon ) &=&{\frac{1}{\pi /2-\epsilon }}\arcsin ({\frac{\cos (\epsilon
)\sinh (x)}{\sqrt{\sinh ^{2}(x)+\sin ^{2}(\epsilon )}}}).  \label{5}
\end{eqnarray}
For fixed non-zero $\epsilon $, each of these is a good function of $x$. The
derivatives are given by 
\begin{eqnarray}
{\frac{dF(x,\epsilon )}{dx}} &=&2{\frac{1-\cos (2\epsilon )\cosh (2x)}{%
[\cosh (2x)-\cos (2\epsilon )]^{2}}}-{\frac{\sin (2\epsilon )}{(\pi
/2-\epsilon )[\cosh (2x)-\cos (2\epsilon )]}},  \nonumber \\
{\frac{dG(x,\epsilon )}{dx}} &=&{\frac{\sin (2\epsilon )}{(\pi /2-\epsilon
)[\cosh (2x)-\cos (2\epsilon )]}}.  \label{6}
\end{eqnarray}
In the limit as $\epsilon \rightarrow 0$, 
\begin{equation}
F(x,\epsilon )\rightarrow {\frac{x}{\left| x\right| }}{\frac{2}{e^{2\left|
x\right| }-1}},\qquad G(x,\epsilon )\rightarrow {\frac{x}{\left| x\right| }}.
\label{7}
\end{equation}
Note that when these limiting values are put in (4), we get exactly the
separation given in Eq. (2) of Ref. 1. What we have done here is to show
explicitly that each term in the separation corresponds to the limit of a
good function.

If we consider the derivatives in this limit, we see that 
\begin{equation}
{\frac{dF(x,\epsilon )}{dx}}\rightarrow -\text{csch}^{2}(x),\qquad {\frac{%
dG(x,\epsilon )}{dx}}\rightarrow 2\delta (x).  \label{8}
\end{equation}
Hence, $dF/dx$ is a good function that goes to $-$csch$^{2}(x)$ for any
finite $x$ and which has the property that its integral from $-\infty $ to $%
+\infty $ is zero. This last follows since $F(x,\epsilon )$ vanishes for $%
x\rightarrow \pm \infty $.

The formula (1) is surprising, since the delta-function at the origin
arises, so to speak, from the behavior at infinity rather than that at the
origin. In this connection it is perhaps worthwhile to make the comparison
with the well known distribution the principal value of $x^{-1}$, which can
be defined as the limit of the good function 
\begin{equation}
P(x,\epsilon )={\frac{x}{x^{2}+\epsilon ^{2}}}.  \label{9}
\end{equation}%
In the limit as $\epsilon \rightarrow 0$, 
\begin{equation}
P(x,\epsilon )\rightarrow P{\frac{1}{x}},  \label{10}
\end{equation}%
where here $P$ denotes the principal value. Like our smooth approximation to 
$\coth (x)$, $P(x,\epsilon )$ is for small $\epsilon $ very close to $1/x$
except in a narrow range about $x=0$, where it turns over and smoothly
connects through the origin. Also, as with our smooth approximation to $%
\coth (x)$, the derivative of $P(x,\epsilon )$ is very close to $-x^{-2}$
except for a narrow range of width of order $\epsilon $ about $x=0$, where
there is a large positive peak. However, in this case the area under the
central peak equals that in the wings, since the net change of $P(x,\epsilon
)$ as $x$ is carried from $-\infty $ to $+\infty $ is zero. Thus, no delta
function appears in the derivative.

Perhaps still more surprising is what we see if we form the difference of
the two functions: $\coth (z)-z^{-1}$, where we have denoted the variable as 
$z$ to emphasize that here we are talking about functions and not
distributions. This difference-function is continuous and bounded for all
real $z$. The same is true of its derivative, so there is no delta function!
What has happened? The answer is that the difference of two functions is not
necessarily the same as the difference of the corresponding distributions.
In this instance, one must take into account the definition of the
distributions at $x=0$, where the functions are undefined. Recall in
particular that csch$^{2}(x)$ in the formula (1) is defined, like the
derivative of $Px^{-1}$, to be a distribution with area zero, so the delta
function must appear.

Figure captions

Fig. 1 The function $\coth (x)$ and its smooth approximation $\coth
(x,\varepsilon ).$

Fig 2. The derivative of $\coth (x)$ and its smooth approximation $\coth
(x,\varepsilon )$.

 Figures are available on request from R.F.O'C.

\begin{references}
\bibitem{note1} This justifies our assertion that in the physical
application described in Ref. 1 this term vanishes in the classical limit.
\end{references}
\end{document}